\documentclass[aps,prl,twocolumn]{revtex4}
\usepackage{amssymb,graphicx}
\newcommand{\nn}{\nonumber}

\newcommand{\fig}[2]{\includegraphics[width=#1]{./figures/#2}}

\newlength{\bilderlength}

\newcommand{\rmd}{{\mathrm{d}}}

\begin{document}
\bibliographystyle{../../macros/KAY.bst}
\title{\sffamily \bfseries \Large 2-loop Functional Renormalization
for elastic manifolds pinned by disorder in $N$ dimensions}
\author{\sffamily\bfseries\normalsize Pierre Le Doussal and Kay
J\"org Wiese \vspace*{3mm}} \affiliation{ CNRS-Laboratoire
de Physique Th{\'e}orique de l'Ecole Normale Sup{\'e}rieure,
24 rue Lhomond, 75005 Paris,  France. }
\date{\small\today}
\begin{abstract}
We study elastic manifolds in a $N$-dimensional random potential using
functional RG. We extend to $N>1$ our previous construction of a field
theory renormalizable to two loops. For isotropic disorder with $O(N)$
symmetry we obtain the fixed point and roughness exponent to next
order in $\epsilon=4-d$, where $d$ is the internal dimension of the
manifold. Extrapolation to the directed polymer limit $d=1$ allows
some handle on the strong coupling phase of the equivalent
$N$-dimensional KPZ growth equation, and eventually suggests an upper
critical dimension $d_{\mathrm{u}}\approx 2.5$.
\end{abstract}
\maketitle

Disordered elastic systems are under extensive study both
theoretically and experimentally. They are of interest for a number of
physical systems, such as CDW \cite{cdw}, flux lattices
\cite{vortices,vortices2}, wetting on disordered substrates
\cite{rolley}, and  magnetic interfaces \cite{creepexp}, where the
interplay between the internal order and the quenched disorder of the
substrate produces pinned phases with non-trivial roughness and glassy
features \cite{book_young}. Typically they are described by elastic
objects, with internal $d$-dimensional coordinate $x$, parameterized
by a $N$-component height, or displacement field $u(x)$. Analytical
methods are scarce, and developing a field-theoretical description
poses a considerable challenge.  One reason is that naive perturbative
methods fail, technically due to the breakdown of the dimensional
reduction phenomenon \cite{dimred2}, and physically because describing
the multiple energy minima in a glass seems to contain some
non-perturbative features. One subset of these problems, the directed
polymer (i.e.\ $d=1$) in a random potential, maps onto the KPZ growth
problem, well known to exhibit a strong coupling phase, which is out
of reach of standard perturbative methods \cite{KPZ}.  It is thus
important to obtain a field-theoretical description of this phase,
since the value and even the existence of its upper critical dimension
is still a matter of considerable debate
\cite{KPZsimu,MarinariPagnaniParisi2000}.

One method which holds promise to tackle this class of problems is the
functional renormalization group (FRG). Although it was introduced
long ago, within a 1-loop Wilson scheme
\cite{fisher,balents_fisher,frgdyn}, it is, not so surprisingly,
hampered with difficulties, and only recently attempts have been made
to push the method further
\cite{balentspld1,balentspld2,twolooplarkin,Scheidl,twoloop,twolooplong,twoloopdep,largeN}. The
main problem is that the effective action at zero temperature becomes
non-analytic at a finite scale, the Larkin scale, where metastability
appears.  Although fixed points are accessible in a $d=4- \epsilon$
expansion, non-analyticity results in apparent ambiguities in the
renormalized perturbation theory at $T=0$ \cite{twoloop,twolooplong}.
These problems are absent at $T>0$ \cite{frgdynT,chauve_pld} (at least
at leading order and for $N=1$) but since temperature is dangerously
irrelevant, the finite temperature description is rather complicated
\cite{balentspld1}.  Until now, it has lead to a complete
first-principle solution of ambiguities (and calculation of the
$\beta$-function to four loop) only for the toy-model limit $d=0$,
$N=1$ \cite{balentspld2}. A case where ambiguities have been resolved
from first principles at $T=0$ to 2-loop order, is the $N=1$ depinning
transition \cite{twoloop,twoloopdep}.  Finally, the FRG has also been
solved in the large-$N$ limit \cite{largeN}. Its solution reproduces,
apparently with no ambiguity, the main results from the
replica-symmetry-breaking saddle point of Ref.~\cite{mezard_parisi},
and also underlies the importance of specifying the system preparation
\cite{largeN}.

In the more difficult case of the statics within the $d=4-\epsilon$
expansion, detailed analysis to two and three loops
\cite{twoloop,twolooplong,3loop} for the case of $N=1$ have suggested
several methods to construct a renormalizable field theory. These
methods give a unique finite $\beta$-function, with non-trivial
anomalous terms. This $\beta$-function satisfies the
potentiality constraint, with anomalous terms distinct from those
 at depinning, and  a fixed point with the same linear
cusp non-analyticity as to one loop, hence confirming the consistency
of the picture.

The aim of this paper is to extend these methods to the $N$-component
model. We show how an extended $\beta$-function can be obtained and
point out the specific features of the case $N>1$.  For the case of
$O(N)$-symmetric disorder we compute the fixed point and roughness
exponent $\zeta$ to next order in $\epsilon=4-d$, where $d$ is the internal
dimension of the manifold. We then study the extrapolations to the
directed polymer limit $d=1$, and discuss the various scenarios for
the strong coupling phase of the equivalent $N$-dimensional KPZ growth
equation. In one of them, a value for the upper critical dimension is
estimated.

\begin{figure}\fig{\columnwidth}{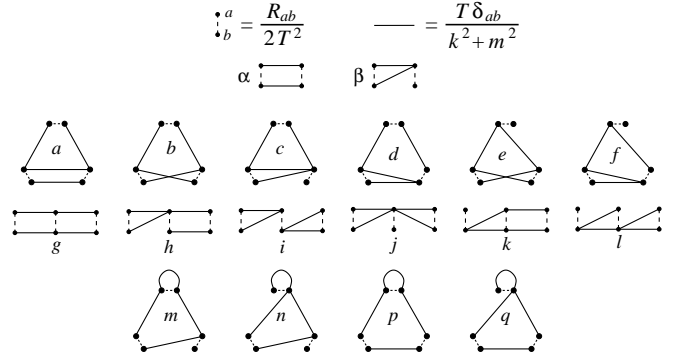}
\caption{Graphical rules, one loop and two loop diagrammatic
corrections to the disorder}
\label{graphs}
\end{figure}

We consider the model for an elastic $N$-component manifold
\begin{eqnarray}
{\cal H} = \int \rmd^d x\, \frac{1}{2} (\nabla u)^2 + V(x,u)
\end{eqnarray}
in a random potential with second cumulant $\overline{V(x,u) V(x',u') }=
\delta^d(x-x') R(u-u')$, where $u=u^i$ is a $N$-component vector.
We derive general equations, and later focus on
the $O(N)$ isotropic case, noting $R(u)=h(r)$ with $r=|u|$.
Introducing replicas we obtain the
replicated action:
\begin{eqnarray}
\frac{{\cal H}_n}{T} = \int \rmd^d x \frac{1}{2T} \sum_{a} (\nabla
u_{a})^2 - \frac{1}{2T^2} \sum_{ab} R(u_a-u_b) \label{H}
\end{eqnarray}
We now carry perturbation theory in the disorder and compute the
one-loop and two-loop corrections to the effective action
$\Gamma[u]$. We use the usual power counting of the $T=0$ theory,
identical to the case $N=1$.  Infrared divergences for $d=4-\epsilon$
only occur in the 2-replica term, which at zero momentum defines the
renormalized disorder; there is no correction to the single replica
term. The graphical rules are depicted in Fig.~\ref{graphs}.  We use
functional diagrams, and mass regularization.  The method and
notations are identical to \cite{twolooplong}, to which we refer for
details.  Here we only stress the differences with the case $N=1$.

The 1-loop correction to disorder (graphs $\alpha$ and $\beta$ in
Fig. 2) reads:
\begin{equation}\label{oneloop}
\delta^{1}R (u) = \left(\frac{1}{2} [\partial_{ij} R(u)]^2 - \partial_{ij}
R(0) \partial_{ij} R(u) \right) I\ .
\end{equation}
Summation over repeated indices is implicit everywhere, and $I=\int_k
G_k^{2}=m^{-\epsilon} \tilde I$ with $G_k=(k^2+m^2)^{-1}$. We define
the dimensionless function $\delta^{1} (R) := m^\epsilon \delta^{1} R$
(recognizable by the parenthesis around the argument $R$).  For later
use we also denote the bilinear form $\delta^{1}(R,R) :=\delta^{1}
(R)$.  This yields the standard 1-loop FRG equations, recalled below,
and $\partial_{ij} R$ develops a cusp non-analyticity at $u=0$ beyond
the Larkin length scale $L_c$. For the $O(N)$ model one has $\partial_{ij}
R=\frac{h'}{r} \delta_{ij} + \hat{u}_i \hat{u}_j (h'' - \frac{h'}{r})
= h''(0) \delta_{ij} + \frac{1}{2} h'''(0) r ( \delta_{ij} + \hat{u}_i
\hat{u}_j) + O(r^2)$ and thus $h'''(0)$ becomes non-zero at $L_c$ ($\hat
u=u/|u|$).

The 2-loop corrections to disorder can be decomposed into a ``normal''
part, which is the complete result when $R(u)$ is analytic
\cite{twolooplarkin}, and an ``anomalous'' part which arises from
non-analyticity. The normal part reads:
\begin{eqnarray} \label{2loop}
&&\!\!\!\! \delta_n^{2} R(u) = (\partial_{ij} R(u) - \partial_{ij} R(0)) \partial_{ikl} R(u)
\partial_{jkl} R(u) I_A \\
&& \!\!\!+ \Big[ \frac{1}{2} \partial_{ijkl} R(u) (\partial_{ik} R(u) {-}
\partial_{ik} R(0)) (\partial_{jl} R(u) {-} \partial_{jl} R(0)) \Big]
I^2 \nonumber.
\end{eqnarray}
The first line stems from diagrams $b$ and $a$ of Fig.\ 1 respectively
and the second from $g,h,i,j$. One has $I_A=\int_{k_1,k_2} G_{k_1}
G_{k_2} G^{2}_{k_1+k_2}=m^{-2 \epsilon} \tilde I_A$ and we denote in
analogy to $\delta^{1} (R)$ the dimensionless function
$\delta^{(2)}(R) := m^{2\epsilon} \delta^{2} R$. The FRG
$\beta$-function is then:
\begin{equation}
- m \partial_m R|_{R_0} = \epsilon [R + \delta^{1} (R)
+ 2 \delta^{2} (R) - \delta^{1,1} (R) ] \ , \label{beta0}
\end{equation}
where the repeated 1-loop counter-term $\delta^{1,1}(R) := 2
\delta^{1}(R,\delta^{1}(R,R))$ arises when reexpressing the bare
disorder $R_0$ in (\ref{H}), in terms of the dimensionless
renormalized one, defined as $m^{\epsilon} R$, as detailed in
\cite{twolooplong}. From (\ref{oneloop}) it reads:
\begin{eqnarray}
\delta^{1,1}(R) &=& [(\partial_{ij} R - \partial_{ij} R(0)) \partial_{ij}
\delta^{1}(R) \nonumber \\
&& ~~- \partial_{ij}
\delta^{1}(R)|_{u=0} \partial_{ij} R ] \tilde I^2 \\
\partial_{ij} \delta^{1}(R) &=& \partial_{ijkl} R (\partial_{kl} R -
\partial_{kl} R(0)) + \partial_{ikl} R \partial_{jkl} R \label{rct}
\end{eqnarray}
The property of renormalizability amounts to cancellation of the
$1/\epsilon$ poles between the two last terms in (\ref{beta0}) using
$\tilde I=\frac{N_{d}}{\epsilon}$ and $\tilde I_A - \frac{1}{2} \tilde
I^2 = N_d^2 (\frac{1}{4 \epsilon} + O (\epsilon^{0}))$ \cite{twoloop}.
The  $\beta$-function (\ref{beta0}) is obtained from
(\ref{oneloop}), (\ref{2loop}) and (\ref{rct}) as:
\begin{eqnarray}
&&\!\!\! - m \partial_m R(u) = \epsilon R(u)
+ \frac{1}{2} [\partial_{ij} R(u)]^2 - \partial_{ij}
R(0) \partial_{ij} R(u)  \nonumber \\
&& + (\partial_{ij} R(u) - \partial_{ij} R(0)) \Big[ \frac{1}{2}
\partial_{ikl} R(u) \partial_{jkl} R(u) - \alpha_{ij} \Big]
\label{frgeq}
\end{eqnarray}
The cancellation works perfectly for the normal
parts. Anomalous parts, to which we turn now,
produce the last term.

We start with the anomalous part of the repeated counter-term:
\begin{eqnarray} \label{mupnu}
 \delta^{1,1}_a (R)= - (\mu_{ij} + \nu_{ij}) \partial_{ij} R(u) \tilde I^2\ ,
\end{eqnarray}
where we denote the limits of small argument $v \to 0$:
\begin{eqnarray} \label{munu} 
\mu_{ij} &:=& \partial_{ikl} R(v) \partial_{jkl} R(v)|_{v \to 0}  \\
\nu_{ij}&:=&  \partial_{ijkl} R(v) (\partial_{kl} R(v) - \partial_{kl}
R(0))|_{v \to 0} 
\end{eqnarray}
which, in general, are direction dependent. For a $O(N)$
model, the third derivative tensor:
\begin{equation}
\partial_{ijk} R(v) = A(r) ( \delta_{ij} \hat{v}_k + \delta_{ik}
\hat{v}_j + \delta_{kj} \hat{v}_i ) + B(r) \hat{v}_i  \hat{v}_j
\hat{v}_k \label{r3}
\end{equation}
with $\hat{v}=v/|v|$, $A(r)=(r h'' - h')/r^2$ and $B(r)= (r^2 h'''
- 3 r h'' + 3 h')/r^2$, has a $\hat{v}$-dependent small $v$ limit
(\ref{r3}) with $A(0)=- B(0)= h'''(0)/2$. This yields:
\begin{equation}
\mu_{ij} = h'''(0)^2 (\frac{1}{2} \delta_{ij}  + \frac{N+1}{4}
\hat v_i \hat v_j) \label{mu}
\end{equation}
and, similarly one finds $\nu_{ij} = \frac{N+1}{4}  h'''(0)^2
(\delta_{ij} - \hat v_i \hat v_j)$.

Let us first superficially examine the structure of the 2-loop 
graphs, following the discussion in \cite{twolooplong}. As for
$N=1$, one can discard $c=d=0$ from parity and similarly set
$m+n=0$ and $p+q=0$. One can then write:
\begin{eqnarray} \label{mupnutilde}
&& \delta^{(2)}_a(R) = - (\tilde \mu_{ij} \tilde I_A + \tilde
\nu_{ij} \tilde I^2) \partial_{ij} R(u)
\end{eqnarray}
where the first term comes from graphs $e$ (more properly, from the
sum of all graphs $a$ to $f$) and the second from graphs $k+l$ (from
the sum of graphs $i$ to $l$). Global cancellation of the $1/\epsilon$
pole in the $\beta$-function works provided $\tilde \mu_{ij} + 2
\tilde \nu_{ij} = \mu_{ij} + \nu_{ij}$. This then produces
$\alpha_{ij}= \tilde \mu_{ij}/2$ in the FRG equation above.

We can now use the methods introduced in \cite{twolooplong} to analyze
the total 2-loop contribution to the effective action, including
possibly ambiguous graphs. One first computes $\Gamma[u]$ in a region
of $u$ where no ambiguity is present, using excluded replica sums, and
constraints valid in the zero-temperature theory (the so-called sloop
elimination method, Section V.B in \cite{twolooplong}). One finds that
extraction of the 2-replica part yields $\tilde \mu_{ij} = \mu_{ij}$
and $2 \tilde \nu_{ij} = \nu_{ij}$, i.e.\ it works as for $N=1$. This
is equivalent to renormalizability diagram by diagram, and thus it
satisfies the global renormalizability condition. The background
method also yields that result (\cite{twolooplong}, Section V.C). The
end result for the $\beta$-function, $\alpha_{ij}=\mu_{ij}$, although
unambiguous for $N=1$, needs further specification for $N>1$, since
the limit in (\ref{mu}) is direction dependent.

Another important consideration for the resulting $\beta$-function
is the issue of the ``super-cusp''. For $N=1$ it was found that
the $\beta$-function is such that the cusp non-analyticity of
$R''(u)$ at $u=0$ does not become worse at two loops. That by
itself constraints the amplitude of the anomalous term, since any
other choice yields a stronger singularity \cite{footnote2}. We now point out that
if $v$ and $u$, in (\ref{mupnu}), (\ref{munu}), (\ref{mupnutilde}),
are colinear, i.e.\ $\mu_{ij}(v)=\mu_{ij}(u)$ then
there is no super-cusp. Indeed the result:
\begin{eqnarray} \label{anomalousfinal} 
\alpha_{ij}(\hat u)= \lim_{r \to 0} \frac{1}{2} \partial_{ikl}
R(r \hat u)
\partial_{jkl} R(r \hat u)
\end{eqnarray}
obviously yields cancellation of the linear term in $u$ in
(\ref{frgeq}) (although it is not the only possibility
\cite{footnote1}). Colinearity of $v=u_{a}-u_{c}$ and $u=u_{a}-u_{b}$
is natural if one computes the
effective action in a background configuration breaking the
rotational symmetry, which appears to be 
required for the present theory to hold.

\begin{figure}
\begin{tabular}{|c||c|c||c|c|}
\hline
$N$ & $\zeta_{1}$ & $\zeta_{1}^{0}$ & $\zeta_{2}$ & 
$\zeta_{2}^{0}$  \\
\hline\hline 1 &0.2082980 & 0.2 &0.0068573 & 0 \\\hline 
2 & 0.1765564 &0.166667 &0.17655636 &-0.00555556 \\\hline
2.5 & 0.1634803 &0.153846 &-0.000417 &-0.00782058 \\\hline
3 &0.1519065  &0.142857 & -0.0029563   &-0.00971817 \\\hline
4.5 & 0.1242642 &0.117647 & -0.009386 &-0.013583 \\\hline
6 & 0.1043517  &0.1 &-0.0135901  & -0.0155556 \\\hline
8 &  0.0856120 &0.0833333 & -0.0162957  & -0.016572 \\\hline
10 & 0.0725621 &0.0714286 &-0.016942  & -0.0166517 \\\hline
12.5 &    0.0610692 & 0.0606061 & -0.0165154 &-0.0161654 \\\hline 
15 &0.0528216  &0.0526316 &-0.01564  &-0.0154217 \\\hline
17.5 &  0.046595 &0.0465116 & -0.0147 &-0.014608 \\ \hline 
20 &0.0417 &0.0416667 &-0.0138   &-0.013804 \\
\hline
\end{tabular}

\fig{.5\columnwidth}{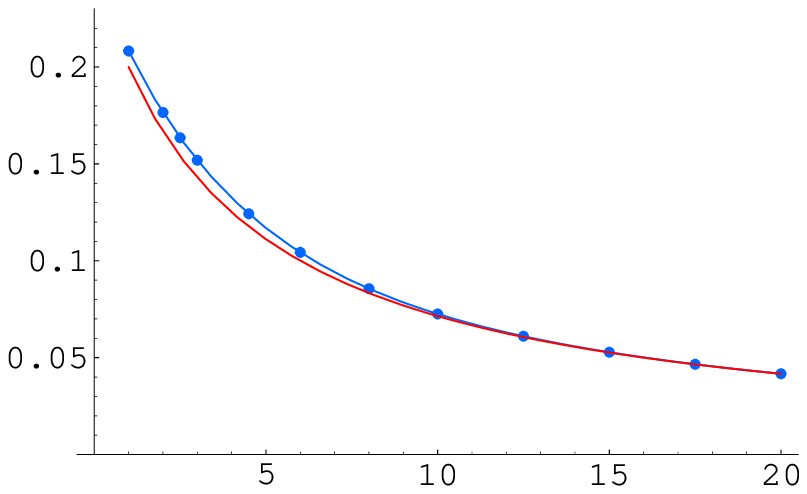}\fig{.5\columnwidth}{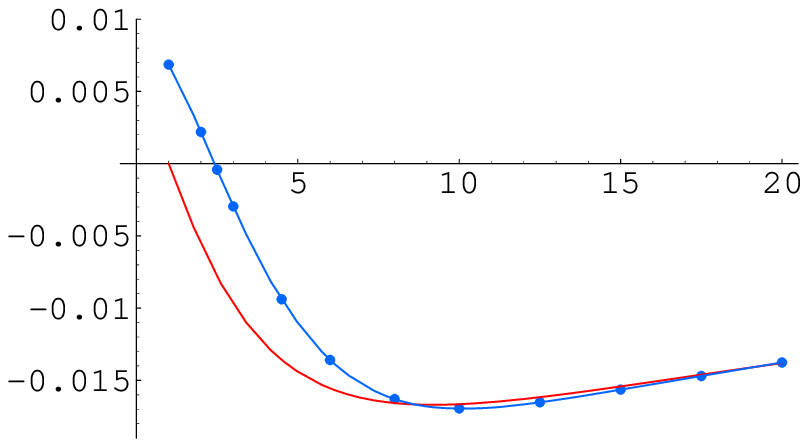}

\caption{Numerical results for the exponents $\zeta_{1}$ and
$\zeta_{2}$ for different values of $N$ (top). Numerical plots of
$\zeta_{1} (N)$ (bottom/left) and $\zeta_{2} (N)$ (bottom/right), in
blue with the numerical values from the table as dots. The
red curves (no dots) represent the asymptotic expansion.}
\label{data}
\end{figure}

We now specialize to the $O(N)$ model. Starting from (\ref{frgeq})
and further rescaling $h(r) \to 
m^{- 4 \zeta} h(r m^{\zeta})$, using $\zeta$
we obtain the following FRG flow-equation to two loops:
\begin{eqnarray}\label{2loopFPE}
-\! &m &\!\partial_m h(r) =
(\epsilon - 4 \zeta) h(r) + \zeta r h'(r) \nonumber \\
&&+\frac{1}{2} h''(r)^2 - h''(0) h''(r)\nonumber \\
&& +\frac{N-1}{2} \frac{h'(r)}{r} \left(\frac{h'(r)}{r} - 2 h''(0)\right)
\nn \\
&&+\frac{1}{2} \left( h''(r) - h''(0) \right) \,{h'''   (r)}^2 \nonumber \\
&&
 +\frac{N{-}1}{2}
\frac{{\left( h'(r) {-} rh''(r) \right) }^2\,
     ( 2 h'(r) {+} r(h''(r) {-}3 h''(0)  )  )}{r^5}
 \nn\\
&& -h'''(0)^{2} \left[\frac{N+3}{8}h''(r)+\frac{N-1}{4}\frac{h'(r)}{r} \right]
\ .
\end{eqnarray}
where the last line arises from the anomalous term (\ref{anomalousfinal}). 
This FRG equation admits for any $N$ a non-trivial attractive fixed
point such that $h''(r)$ has a linear cusp at the origin and decays to
0 at infinity faster than a power law, thus corresponding to short range (SR)
disorder. Finding the associated $\zeta
= \zeta_1  \epsilon + \zeta_2 \epsilon^2 + O(\epsilon^3)$ is an eigenvalue problem,
which has to be solved order by order in $\epsilon$ following
\cite{twoloop,twoloopdep,twolooplong}.  Our results for $\zeta_1$ and
$\zeta_2$ are given on Fig.\ \ref{data}.
Although for SR disorder no analytical expression can be found for
$\zeta_1$ and $\zeta_2$, their large-$N$ behavior can be obtained
from an asymptotic analysis of (\ref{2loopFPE}). Let us extend the
analysis of Balents and Fisher (BF) \cite{balents_fisher}. Define
$h=1/N \hat h$, $y=r^2/2$ and $\hat h(r)=Q(y)$. For $y \gg 1$ the FRG
equation can be linearized:
\begin{equation}\label{linear}
(\epsilon - 2 \zeta) Q' + 2 \zeta y Q'' - (A + 3 B) Q'' - 2 B y Q''' = 0 
\end{equation}
with $A=(1 - \frac{1}{N} ) Q'_0 + \frac{N-1}{4 N^2} \hat h'''(0)^2$
and $B = \frac{1}{N} Q'_0 + \frac{N+3}{8 N^2} \hat h'''(0)^2$, $\hat
h''(0)=Q'(0)=Q'_0$. BF noted that there is an overlapping region $1
\gg y \gg N$ where the solution can also be found perturbatively by
expansion in $1/N$, yielding for $Q$ a pure exponential. It is indeed
an exact solution of (\ref{linear}), with a unique value for
$\zeta_1$, the BF result $\zeta_1 \approx \zeta_1^0$ with $\zeta_1^0 =
1/(4+N)$ (i.e.\ the result from the replica variational method
\cite{mezard_parisi}).  The corrections (which arise from the
neglected non-linear terms) are shown to be exponentially small; a
more accurate estimate being $\zeta_1 \approx \zeta_1^1$ with
$\zeta_1^1=\zeta_1^0+ (N+2)^2/(N+4)^2 2^{-(N+2)/2}/(4 e)$. To next
order we find similarly the approximation to $\zeta_2 $ \cite{footnoteeps}:
\begin{eqnarray}
 \zeta_2^0 = - \frac{(N^2-1)(2+N)}{2 (4+N)^3 (3+N)} \ ,
\end{eqnarray}
\begin{figure}[b]
\centerline{{\unitlength0.95mm
\begin{picture} (90,55)
\put(1,0){\fig{85\unitlength}{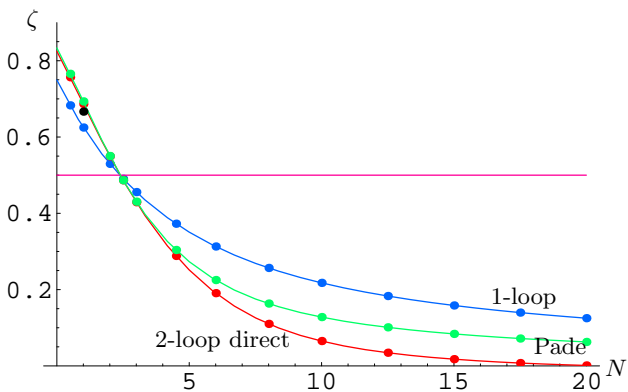}}
\put(5,52){$\zeta $}
\put(86,3){$N$}
\put(70,13){1-loop}
\put(76,6){Pade}
\put(23,7){2-loop direct}
\end{picture}}
} \caption{Results for the roughness $\zeta$ at 1- and 2-loop order,
as a function of the number of components $N$. We both show a direct
extrapolation and the Pade (1,1): $\zeta_{\mathrm{Pade}}=\frac{\epsilon
\zeta_{1}}{1-\epsilon \zeta_{2}/\zeta_{1}}$.}  \label{f:Ncomp}
\end{figure}%
where we have not attempted to estimate further
corrections, presumably again exponentially small at large $N$.  We
note that $\zeta_2^0$ arises {\it from the anomalous terms
only}. These estimates are listed and plotted on Fig.\ \ref{data}
together with the numerical solution of (\ref{2loopFPE}). The quality
of the large-$N$ analysis is quite remarkable.

We now discuss the extrapolation of our result to the directed polymer
(DP) case $d=1$, $\epsilon =3$, plotted in Fig.\ \ref{f:Ncomp}. We see
that the 2-loop corrections are rather big at large $N$, so
extrapolation down to $\epsilon=3$ is difficult. However both 1- and
2-loop results as well as the Pade-(1,1) reproduce well the two known
points on the curve: $\zeta =2/3$ for $N=1$ \cite{KPZ} and $\zeta =0$
for $N=\infty$ \cite{largeN}. This branch in Fig.\ \ref{f:Ncomp}
corresponds to zero temperature and a continuum model. On the other
hand we find that for all curves in figure \ref{f:Ncomp} the roughness
$\zeta $ becomes smaller than the thermal
$\zeta_{\mathrm{th}}=\frac{1}{2}$ at $N=N_{\mathrm{uc}}\approx
2.5$. This naturally suggests the scenario that at non-zero
temperature $\zeta=1/2$ for $N\ge N_{\mathrm{uc}}$, i.e.
$N_{\mathrm{uc}}$ is the upper critical dimension \cite{vortices}. The
same argument gives an upper critical dimension $N_{\mathrm{uc}}$ for
the KPZ-equation of non-linear surface growth \cite{KPZ,Laessig}. On the other
hand, simulations on discretized models of both the directed polymer
(at $T=0$) and the KPZ equation
\cite{KPZsimu,MarinariPagnaniParisi2000} suggest that $\zeta > 1/2$ in
all dimensions, but should be taken with caution \cite{footKPZ}.
Since the FRG is a systematic expansion in $\epsilon =4-d$, such a
scenario seems reconcilable with our above results only through
non-perturbative corrections in $\epsilon$, possibly non-analytic at
$\epsilon =2$.

To conclude we have obtained for the $N$-component model a FRG
description at 2-loop order. Various studies, including at large $N$,
are under way to obtain a better understanding of the structure of the
theory. For the KPZ growth and the directed polymer we have improved
the determination of the possible upper critical dimension. Further
numerics, in particular for the directed polymer at $T=0$ would be helpful.



\begin{thebibliography}{999}
\bibitem{footnote1}
For the $O(N)$ model one must have $\alpha_{ij}= \alpha
\delta_{ij} + \beta \hat u_i \hat u_j$ and the absence of
super-cusp reads $\alpha (1+N) + 2 \beta = \frac{1+N}{2}
h'''(0)^2$.
\bibitem{footnote2}
An alternative scenario is that a non-linear
cusp with dimension-dependent exponent develops, e.g. $h''(r) \sim
r^{1+A \epsilon}$. It may require deviations from
the simplest scenario of a uniform density 
of shocks of codimension one.
\bibitem{footKPZ}
Note that the UV-cutoff of the RSOS model in
\cite{MarinariPagnaniParisi2000} is in the middle of the scaling
plots for high dimensions.
\end{thebibliography}

\end{document}